# Sizing of Energy Storage System for Virtual Inertia Emulation

Mohamed Abuagreb
Electrical and Computer Engineering
Clemson University
Clemson, SC, USA
Mobuaigre@clemsen.edu

Ahmed Abuhussein
Electrical and Computer Engineering
Gannon University
Erie, PA, USA
abuhusse001@gannon.edu

Saif alZahir
Electrical and Computer Engineering
University of Concordia
Montreal, QC, Canada
saifz@uvic.ca

*Abstract*— The infusion of renewable energy sources into the conventional synchronous generation system decreases the overall system inertia and negatively impacts the stability of its primary frequency response. The lowered inertia is due to the absence of inertia in some of the renewable energy-based systems. To maintain the stability of the system, we need to keep the frequency in the permissible limits and maintain low rotational inertia. Some authors in the literature have used the virtual synchronous generators (VSG) as a solution to this problem. Although the VSG based distributed recourses (DER) exhibits the characteristics and behavior of synchronous generators (SG) such as inertia, frequency droop functions and damping but it does not optimally solve the question of frequency stability. This paper presents a solution for these problems via an empirical model that sizes the Battery Energy Storage System (BESS) required for the inertia emulation and damping control. The tested system consists of a Photovoltaic (PV) based VSG that is connected to a 9-Bus grid and the simulation experiments are carried out using EMTP software. The VSG transient response is initiated by a symmetric fault on the grid side. Our simulations show the battery energy sizing required to emulate the virtual inertia corresponding to several design parameters, i.e., the droop gain, $K_\omega$, the droop coefficient, $K_d$, and the VSG time constant $T_a$. The simulation results show that to limit the rate of change of frequency (ROCOF) the battery needs to absorb a peak power of 0.57 pu and supply a peak power of 0.63 pu when the time constant is 4s, and the battery needs to absorb a peak power of 0.62 pu and supply a peak power of 0.69 pu when the time constant is 10s. Moreover, the findings suggest that to achieve a better ROCOF, $K_d$ and $T_a$ can be increased, however, this requires a larger battery rating. A better ROCOF can also be achieved by increasing the droop gain, $K_\omega$, without increasing the battery size.

## I. Introduction

Recently, a large amount of renewable energy sources (RESs), generally photovoltaic (PV) and wind power generation plants, have been introduced and connected to national power grids worldwide. Such combination has created new technical challenges that, if not attended to, will compromise the stability of the power system. It has been shown in the literature that the increased penetration of RESs may significantly reduce the system inertia as RESs have some inadequacies in providing fundamental inertial and primary frequency control (PFC) response [1], [2], [3]. In addition, the rate of change of frequency (ROCOF) and the max/min frequency limits during post-contingency periods will be affected, and therefore, the provision of frequency control becomes a challenging undertaking to address [3]. Conventional synchronous generators have been adopted by many researchers to provide inertial response to frequency deviation and to compensate for frequency control [3], [4], [5]. Although this approach is effective but it is not sufficient.

One possible solution that has been proposed and implemented to improve the dynamic response of such grids is the provision of additional inertia, albeit virtually [2]. Virtual inertia can be established in distributed generation (DG) by incorporating energy storage with appropriate control mechanisms for the converter. This arrangement will provide a tool to emulate the behavior of a conventional synchronous generator via controlling the output of the converter. In such a way, the DG can be controlled by the converter to exhibit responses similar to that of a real synchronous machine when there are changes in the operating conditions or when disturbances occur in the power grid. This concept is generally known as a virtual synchronous generator (VSG) [4], [5]. Several researchers showed that the VSG can provide frequency droop in the grid and the damper supports the grid oscillation adequately [3], [4], [5], [6]. Based on the above, the renewable energy source can provide ancillary services to the power grid through this means [2].

This paper considered the system models in [4] and [5] and presents methods for ratings battery power and energy for the emulated inertia application. Traditionally, the electric power networks are operating at a normal steady state where the total generated power meets the total load plus system losses and the frequency experiences small, slow variations in response to changes in load or renewable generation output. During abnormal operating conditions, the system's frequency will exhibit larger oscillatory variations. This paper will provide a remedy to such cases.

## II. System Description

The PV system in the model shown in the Fig. 1 is connected to a small power grid through a step-up transformer. The total load connected to the system is approximately 16 MW. The PV supplies a steady-state output active power of 2.75 MW. The remaining power is supplied by the grid source(s). The system initially operates in steady-state. To evaluate the ability of the combined PV and energy storage system on the system response disturbances were created applying three phase faults to some critical transmission lines in the grid with the fault cleared after 200 msec. These events will disconnect some loads from the power grid momentarily. For the period of time under consideration, the PV is generating more power than required, hence resulting into generation-load imbalance in the power system. As shown in Fig.1, the battery energy storage system is combined to the grid-tied inverter to enhance the frequency control and power stability of the PV/BESS system. The implemented model for the VSG is based on the concept of dual modeling of the SG and they are represented in the same way a as in [3].

The outer control loop of the conventional SG is used. Instead of the traditional phase locked loop (PLL), the VSG control implemented in this paper utilizes swing equation to synchronize with the grid, thus the built-in droop and inertia characteristics of the VSG control can provide better active power-frequency response for grid-connected converters [3] and [4]. Different operating conditions can be accommodated by adjusting the inertia constant and damping ratio of the VSG controls to damp any power system oscillation [5].

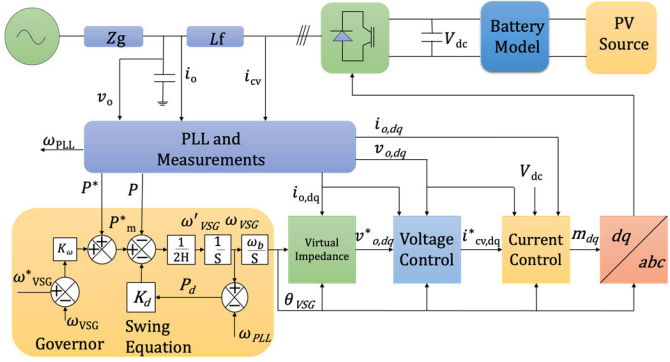

Fig. 1. Proposed Active Power control loop of VSG emulation based on swing equation

The application of the VSG studied in this paper is based on the conventional swing equation, which mimics the inertia in the SG. The implementation of the swing equation is linearized in relation to the speed as shown in equation (1). The virtual inertia's acceleration is calculated by power balance as represented in the equation:

$$\frac{d\omega_{VSG}}{dt} = \frac{1}{T_a}(P_m^* - P_e) - P_d \qquad (1)$$

Where:

$P_m^*$ is the pu virtual machine power input. $P_e$ is active power electrical measured from the grid. $P_d$ is the damping power of VSG. $T_a$ is the time constant of virtual inertia which represents the inertia constant 2H in the conventional SG. As represented in the block diagram in the Fig. 1, the angular speed of the VSG is defined $\omega_{VSG}$ which is integral from derivative of angular speed, while $\theta_{VSG}$ is resulting from the integral of the $\omega_{VSG}$. Also it describes the model of VSG based on the conventional swing equation. The damping power $P_d$ mimics the power damping impact in the conventional SG, which implemented in the virtual synchronous generator based on the difference between $\omega_{VSG}$ and the measured PLL speed $\omega_{PLL}$. The resulting difference is multiplied by the damping factor of VSG, $K_d$ [3] and [4].

In addition, part of Fig.1 shows the droop control representing the equivalent frequency by the governor of VSG at steady state characteristic. The virtual mechanical power, $P_m^*$, in the external loop is used as the reference input for the VSG. Thus, the active power control is obtained by multiplying the droop gain, $K_\omega$ with the difference between the reference VSG speed, $\omega^*_{VSG}$ and actual speed $\omega_{VSG}$ and finally by adding to the external active power reference, $P^*$ [4].

III. BATTERY ENERGY STORAGE SIZING

Owing to the rapid introduction of renewable energy in power generation, the frequency instability in the power system may become a major concern due to the reduced inertia. The swing rated for the response due to a large change in load or generation can exceed the ability of convention control mechanisms to respond to stabilize the system. Energy storage systems can be used to emulate the response of large synchronous machines [4]. This research proposes adding energy storage on the dc link of PV inverters to provide inertia emulation. Ignoring the power losses, the power balanced between the PV generation, power from the storage system and the inverter output power is given in the equation (2) below:

$$P_{load} = P_{PV} + P_{BESS} \qquad (2)$$

Where $P_{PV}$ is the Photovoltaic power and the $P_{BESS}$ is the power absorbed or supplied by the storage energy system. The VSG model described above controlled the real power set point for the inverter based on the swing equation shown in Fig. 1. The energy storage connected to the dc bus of the inverter enabled this swing response. There are two methods to adjust the inertia response when severe disturbance occurs in the PV generation as proposed in the [4]. In the first method, the frequency reserve capacities that can be increased by increasing the responsive reserve generation, usually in the form of gas turbines. In addition, accommodating the alternating nature of the PV generation results in frequency ramping of conventional generation, the power plant components' efficiency. Therefore, the coordination between ESS and PV control is an important requirement, which is employed in this research to enhance the output power of the PV integration and support the frequency. In addition, when the frequency becomes unstable, the BESS helps to enhance system stability by regulating the frequency of the power system, which is similar to the synchronous machine's inertia behavior. To create PV output power that would be similar to that of the SG output power under the principle of frequency regulation standards, new characteristic control converter of PV generation storage was created by adding the VSG technology. In most designs of the VSG, PV generation is combined with BESS and controlled on the dc bus of the converter.

In the second method, the output of the PV generation is operated below the maximum power point with the generation level derated [5]. The additional capacity, which is analogous to the spinning reserve in synchronous generator, supplies the required inertia response when there is disturbance in the system. The downside of this method is the lost revenue due to operating below the maximum power point. This approach is only economical if there is a market for supplying ancillary services.

The energy storage required to support the system with low rotating inertia due to combine of large amount of the PV generation and estimate size these devices to keep stability in the system. To maintain stability in the power system, some researchers proposed sizing of the battery energy storage system devices is to be about 10% of the distributed generation capacity [7]. The steady power transfer from a synchronous machine can be approximated by equation (3) if the stator resistance is neglected. Where, $E_a$ is the armature voltage, $E_t$ is the terminal voltage, $X_t$ is the equivalent reactance, and $\delta$ is the load angle.

$$P_e = \frac{E_a.E_t}{X_T}.sin\delta \qquad (3)$$

The output from the PV stays constant during the power system disturbance. The energy storage supplies or absorbs power to allow the inverter to be controlled as a VSG. Equation (4) shows the role of the energy storage in the virtual synchronous generator control [7] and [8].

$$\frac{2H}{\omega_s}.\frac{d\omega}{dt} = P_m^* - P_e \pm P_{Bess} \qquad (4)$$

Where, H is the inertia constant kg.m² and $P_m^*$ is the virtual machine input power.

If the change in angle of the synchronous machine in response to a disturbance is too large, the machine may lose synchronism with the ac system. Otherwise, If the machine has more inertia, the change in angle will be smaller and the machine will be likely to lose stability. In the case of the VSG, the energy storage system is sized to emulate the behavior of the generator. To simplify the calculation for rating the energy storage the small signal response of the VSG around the operating point is linearized and the emulated mechanical input power is assumed to be constant [9] and [7].

The battery energy capacity is calculated based on (5), Where $ESS_d$ is the total energy delivered and $ESS_s$ is the total energy stored (absorbed). The total energy supplied and absorbed is equal to the sum of power deviation over the power swing period (*T*). Ps is the power stored (negative swings) and Pd is the power delivered (positive swings).

$$E_{batt} = ESS_s + ESS_d = \int P_s(t).dt + \int P_d(t).dt = \int_T \Delta P_{batt}.dt \qquad (5)$$

IV. SIMULATION AND DISCUSSION

To validate the utilized model, two scenarios were considered each of which include two cases. The results of these cases are presented in this section below. During the simulations, the inertia constant, damping coefficient and frequency droop settings were varied in order to empirically find a preferred battery size to ensure PV adequate power system stability. Two different inertia time constants were considered for both scenarios. For each case in the scenario, two droop gain ($K_\omega$) values were considered while keeping the damping factor ($K_d$) constant. In both cases, the irradiation of the PV is assumed to be constant providing 2.75 MW. The scenarios are implemented for sizing the battery real power and energy ratings that satisfy the desired frequency deviation metrics.

There are two test scenarios; the first the inertia time constant is set to 4 seconds. In the second scenario the time constant is set to 10 seconds. For each scenario the response is shown with the damping coefficient enabled (case 1 $K_d=400$) and with the damping *disabled* (case 2 $K_d=0$). In each case, the droop gain, $K_\omega$, is set to 20 pu and 40 pu respectively.

A. *Scenario 1: $T_a = 4$*

1) *Case 1: $K_d = 400$*

In this case, the effect of virtual synchronous generator control parameters on the damping coefficient and frequency droop characteristic, are examined and tested. This test was conducted using damping coefficient $K_d = 400$ while we used two droop gains $K_\omega = 20$ pu and 40 pu, respectively. The impact of such assumption was calculated in terms of inertia response to the system disturbance. We observed that an instantaneous loss of power and a change in frequency dynamics occur when a generator is dropped. Since both of these characteristics are time varying, the worst-case is considered for analysis. Theoretically, the use of frequency deviation in the swing equation response due to a system disturbance can be used to estimate the required stored energy.

The proposed method is computationally simple and provides a rapid estimate of the frequency response, and later, ESS sizing. As the contingency size depends upon the system settings, the event chosen in this research was the loss of 2.749 MW for 0.2 sec, where the ESS provides upwards frequency support in the case of generator outage and downwards support in the case of load outage.

The active power response at the point of interconnection is shown in Fig. 2. After the event, the energy storage system capacity and control settings are supposed to improve the frequency response of the grid.

Fig. 3 shows the frequency deviation at the point of interconnection with the energy storage system that is controlled for the two values of inertia time constants. The resulting frequency deviation was 0.5 Hz for $P_{PV} = 2.75$ MW at an inertia time constant $T_a = 4$ sec. The rate of change of frequency at the PCC connection is important to prevent mis-operation of distance protection relays and must not be higher than 0.5 Hz/sec.

The virtual synchronous generator converts the delivered power from the PV generation combined with the energy from the battery to control the power that will be supplied to the loads through the power network while emulating the behavior of a synchronous generator.

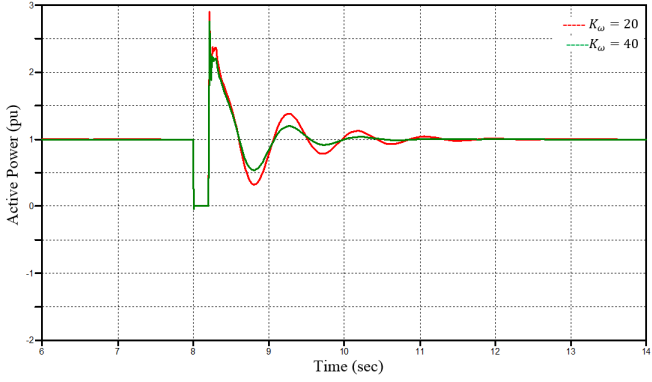

Fig. 2 Comparing power response at PCC with $T_a = 4$ sec and $K_d = 400$ following a fault

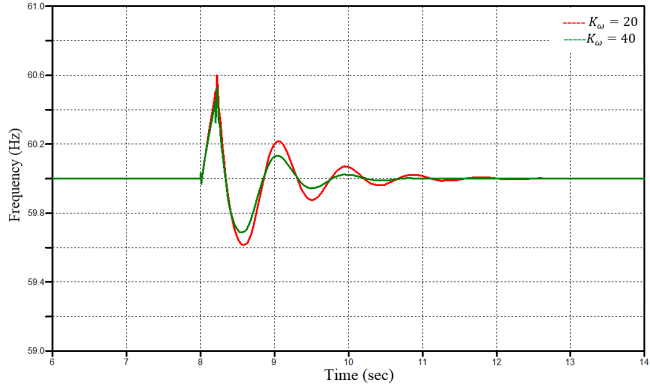

Fig. 3 Comparison of the system frequency response to disturbances with $T_a = 4$ sec and $k_d = 400$ at different droop gains

*2) Case 2: $K_d = 0$*

Based on control system theory the damping coefficient is an important parameter for system stability. A similar concept is used for the virtual synchronous generator to create damping effect for the system stability. To verify the damping effect in the model, $K_d$ was decreased to observe the system response. In this case, the equation of motion of the rotor was applied without damping. The contingency created for case 1 is repeated in case 2 with the damping gain $K_d$ set to zero, and $K_\omega = 20$ pu and 40 pu respectively. As a result, the frequency swing was very high at point of interconnection due to the absence of the damping component as shown in Fig. 4. This demonstrates that the damping component is required to damp the oscillation.

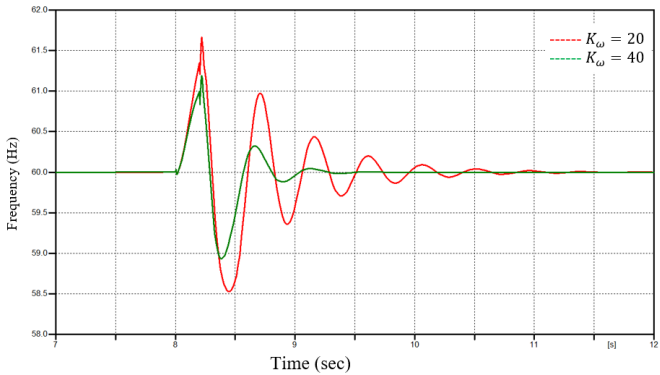

Fig. 4. Comparison of the system frequency response to disturbances with $T_a = 4$ sec and $K_d = 0$ when $K_\omega$ is varied

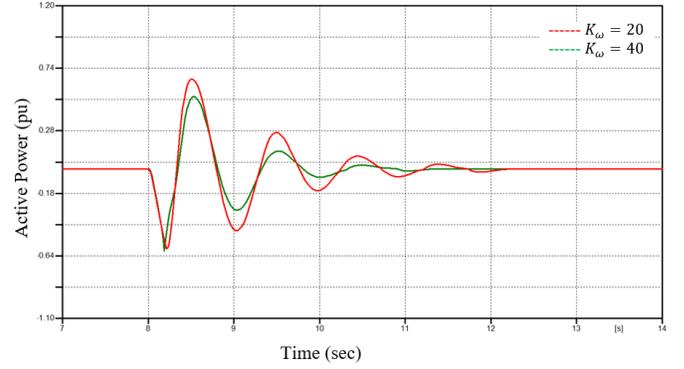

Fig. 5. Comparison of the power response at PCC with $T_a = 4$ sec and $K_d = 0$ when $K_\omega$ is varied

In this case, the high gain of $K_\omega$ lead to less deviations of the frequency and a lower rate of change of frequency and battery discharging.

The swing equation is used to calculate the power delivered by energy storage system by integrating the area under the power curve between the start of the event and the first zero-crossing. The output power from the battery for the two values of the droop gain due to the generator contingency are shown in Fig. 5. The battery output changes due to the disturbance event that caused frequency deviation in the power system. By comparing the applied droop gain $K_\omega = 20$ and $K_\omega = 40$, it can be concluded that the higher gain results in more damping of the system frequency. However, due to ROCOF limitation, excessive high gain will cause the dynamic response to deteriorate and cause the control system to be unstable.

The comparison of the response obtained evaluating the VSG response due to the event, the power required for each case is described in Table I. The ROCOF, $df/dt$, $F_{max}$ and $F_{min}$ are presented in this table. After performing all tests, in order to provide the inertia response in term of ROCOF as in test 1, it was found that the most severe contingency occurs when $K_d$ is zero and the $K_\omega$ is 20 pu and requiring the largest peak power rating for the battery. Moreover, with the removal of the damping $K_d$, the system frequency and ROCOF increases considerably. The increase of the frequency response can cause the system to be unstable. The ROCOF values and frequency deviation at the PCC increases when the $K_d = 0$ for $K_\omega = 20$ and decrease a little bit when $K_\omega$ is increased to 40 pu. However, when implemented with $K_d$ in the system, the ROCOF with $K_\omega = 20$ pu decreases and increases with $K_\omega = 40$ pu. With small inertia $T_a = 4$ sec, the ROCOF and frequency deviation is still at higher value. The battery power required in test scenario 1 is (0.57 pu) during charging and (0.63 pu) during discharge.

Table I: Frequency deviation, battery power measured for test scenario 1 $T_a = 4s$

| Case | | $F_{min}$ | $F_{max}$ | $df/dt$ | Battery Power (pu) | | |
| --- | --- | --- | --- | --- | --- | --- | --- |
| $k_d$ | $K_\omega$ | | | | Charge | Discharge | Range |
| 400 | 20 | 59.65 | 60.5 | 0.26 | 0.57 | 0.63 | 1.2 |
| | 40 | 59.7 | 60.4 | 0.25 | 0.50 | 0.52 | 1.02 |
| 0 | 20 | 58.5 | 61.5 | 0.57 | 0.51 | 0.74 | 1.25 |
| | 40 | 59.0 | 60.8 | 0.40 | 0.47 | 0.64 | 0.93 |

In order to decrease the frequency deviations and rate of change of frequency with the limit values which is less than 0.5 Hz/sec, we have to increase the $T_a$ to limit the ROCOF value from exceeding $\geqslant 0.5$ Hz/sec and try to find the battery size to provide sufficient system inertia.

*B. Scenario 2: $T_a = 10s$*

In this scenario, the inertia time constant is set to 10 seconds. Below are the two cases considered in this scenario:

   *1) Case 1: $K_d = 400$*

The test in scenario 1 event was repeated using the higher inertia value to observe the system response and determine battery size. This test was conducted to assess the impact of increasing the virtual inertia time constant of VSG.

The system experiences a grid frequency change similar to that observed in case 1 under the first test scenario. The parameters of $K_d$ and $K_\omega$ implemented in this case are similar to that considered in the test scenario 1. The damping coefficient $K_d$ in this case is set to 400, while $K_\omega = 20$ pu and 40 pu. Fig. 6 shows the response of the frequency deviation at the point of interconnections. It shows the comparison of the system response to different droop frequency gains ($K_\omega = 40$ pu and $K_\omega = 20$ pu). The system returns to nominal frequency faster when $K_\omega = 40$ pu compared to $K_\omega = 20$ pu.

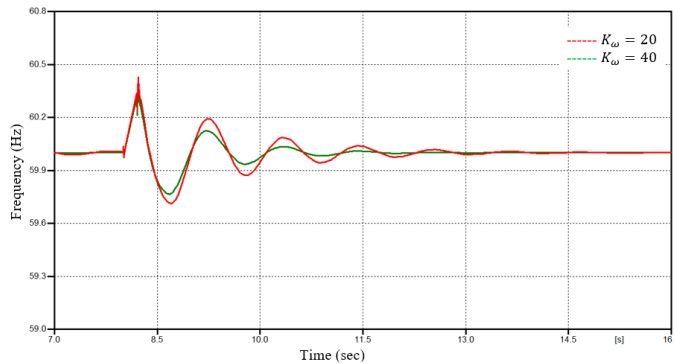

Fig 6. Comparison of the system frequency response to disturbances with Ta= 10 sec and $K_d$ =400 when $K_\omega$ is varied

   *2) Case 2: $K_d = 0$*

Test is repeated with the damping coefficient $K_d$ in set to 0 in this case, while the inertia time constant and the droop frequency gains are kept constant.

As it is observed in Fig. 7, the response of the frequency at point of interconnection tends to last for longer time compared to case 1. In addition, the figure shows that frequency at PCC experiences less oscillation when the $K_\omega = 40$ pu compared to the case with a $K_\omega$ of 20 pu.

The simulation results from test 2 are summarized in the Table II. The increase in the virtual inertia leads to reduced frequency overshoot and provides slower frequency response. As explained before, when big disturbances occur, the rotor of the conventional synchronous generator accelerates or decelerates to enhance stability of the system. In the virtual synchronous generator, the battery energy storage system works working in a similar manner.

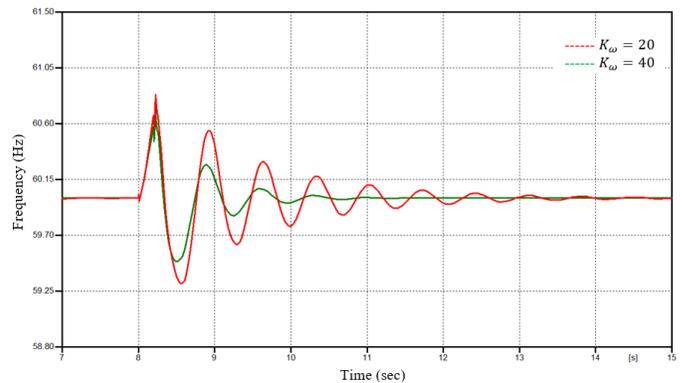

Fig 7. Comparison of the system frequency response to disturbances with Ta= 10 sec and $K_d$ =0 when $K_\omega$ is varied

As illustrated in the Tables I and II, the power system stability depends on the combined response of the components in the system. The rate of change of the power ratio to the ROCOF is directly dependent on the system inertia. A system with larger total inertia can withstand larger disturbances when compared to a system with small inertia. Furthermore, the small inertia system may see large ROCOF value for the large disturbance which may cause nuisance tripping of protective relays. The reason behind this is that large inertia system has slower ROCOF as compared to small inertia system. The virtual synchronous generator needs increased energy storage capacity in order to emulate a larger inertia.

Table II: Frequency deviation, battery power measured for test scenario 1 $T_a$=10s

| Case | | $F_{min}$ | $F_{max}$ | df/dt | Battery Power (pu) | | |
| --- | --- | --- | --- | --- | --- | --- | --- |
| $k_d$ | $K_\omega$ | | | | Charge | Discharge | Range |
| 400 | 20 | 59.7 | 60.32 | 0.116 | 0.62 | 0.67 | 1.29 |
|  | 40 | 59.75 | 60.3 | 0.112 | 0.59 | 0.57 | 1.16 |
| 0 | 20 | 59.3 | 60.8 | 0.37 | 0.57 | 0.69 | 1.26 |
|  | 40 | 59.45 | 60.6 | 0.35 | 0.48 | 0.53 | 1.01 |

In order to fulfill all operational requirements in these tests with high impact disturbances, it can be concluded from the results that the battery power needed in the test scenario 2 is 0.62 pu during charging and 0.69 pu during discharge. This section presents the results for the vary limited test conditions that were considered in order to find a battery peak power rated size (MW) that meet the project goal.

The objective of this study is to investigate the change of inertia coefficient gain and frequency droop gains needed to allow a significant renewable energy source integration and determine the battery power rating.

## V. CONCLUSION

Virtual synchronous generator is a technique used to add inertia to inertia-less or low inertia power electronics-based renewables. The advantage of the emulated inertia is that it can damp the renewable energy source transient response. Thus, ensuring stability after grid disturbances. The emulated inertia (VSG) is achieved by adding an energy storage to the PV Plant. This paper targeted the following issues: 1) to empirically estimate the battery size required to emulate the virtual inertia via simulation as well as the area under the curve of the power swing; 2) to provide a guideline on selecting the virtual governor and the swing equations parameters and show their impact on sizing the BESS. The simulation results illustrated the possible impacts of high renewable energy integration and proposed a simple technique for approximating battery sizing with respect to power and energy by providing emulation inertia to meet the target system, power and frequency characteristics. Two test scenarios were conducted. Sscenario 1 is performed when the inertia added is low $T_a$ = 4s and Scenario 2 is performed when the inertia was high where $T_a$ = 10s. With each scenario, the droop gain and the damping coefficient are varied to observe their impact on the ROCOF and the battery size required to achieve it. Test scenario 1 show that the battery needs to absorb a peak power of 0.57 pu and supply a peak power of 0.63 pu. The results of scenario 2 show that the battery needs to absorb a peak power of 0.62 pu and supply a peak power of 0.69 pu.